\documentclass[aps,prd, onecolumn,groupedaddress,10pt]{revtex4}
\usepackage{amsmath,empheq,color}
\usepackage{amsfonts}
\usepackage{amsthm}
\usepackage{amssymb}
\usepackage{graphicx}
\usepackage{hyperref}
\usepackage[normalem]{ulem}
\usepackage{soul}

	\newcommand{\ncd}{\newcommand}
	\ncd{\mrm}    {\mathrm}
	\ncd{\beq} {\begin{equation}}
	\ncd{\eeq} {\end{equation}}

\begin{document}

\title{Exact Baker-Campbell-Hausdorff formula for the contact Heisenberg algebra}
\author{Alessandro Bravetti}
 \email{alessandro.bravetti@iimas.unam.mx}
 \affiliation{Instituto de Investigaciones en Matem\'aticas Aplicadas y en Sistemas, Universidad Nacional Aut\'onoma de M\'exico,
	Ciudad Universitaria, Ciudad de M\'exico 04510, Mexico}

\author{Angel Garcia-Chung}
\email{angel.garcia@correo.nucleares.unam.mx} 
\affiliation{Instituto de Ciencias Nucleares, Universidad Nacional Aut\'onoma de M\'exico, Ciudad Universitaria, Ciudad de M\'exico 04510, Mexico}

\author{Diego Tapias}
\email{diego.tapias@nucleares.unam.mx}
\affiliation{Departamento de F\'isica, Facultad de Ciencias, Universidad Nacional Aut\'onoma de M\'exico, Ciudad Universitaria, Ciudad de M\'exico 04510, Mexico}

\date{\today}

\begin{abstract}
In this work we introduce the contact Heisenberg algebra which is the restriction of the Jacobi algebra on contact manifolds to the linear and constant functions. 
We give the exact expression of its corresponding Baker-Campbell-Hausdorff formula. We argue that this result is relevant to the quantization of contact systems.
\end{abstract}

\maketitle


\section{Introduction}
The Baker-Campbell-Hausdorff (BCH) formula is an exact result for the calculation of $Z(X,Y)={\rm ln}({\rm e}^{X}{\rm e}^{Y})$ whenever $X$ and $Y$ are elements of a Lie algebra. Explicitly, the BCH formula~\cite{achilles2012early, van2015special, hall2015lie} reads
	\begin{equation}\label{BCH1}
	Z(X,Y) = X + Y - \int^1_0 dt \, \sum^{+\infty}_{n=1} \frac{\left(I - {\rm e}^{L_X} {\rm e}^{t L_Y}\right)^n}{n(n+1)} Y\,, 
	\end{equation}
where $L_X Y := [X,Y]$ and the exponential ${\rm e}^{L_X}$ is defined as
	\begin{equation}
	{\rm e}^{L_X} Y := \left[ I + L_X + \frac{1}{2!} L_X L_X + \cdots \right]Y = Y + [X,Y] + \frac{1}{2!} [X, [X,Y]]+ \dots \label{ExpConm}
	\end{equation} 
	
{In the case of finite dimensional Lie algebras, the expression (\ref{BCH1}) provides an explicit formula to compute $Z(X,Y)$ given only the knowledge of the commutators of the Lie algebra. Therefore, it is of interest for several areas of mathematics and physics. For instance, the theory of Lie groups and Lie algebras, linear partial differential equations, numerical analysis, control theory, sub-Riemannian geometry, and in quantum and statistical mechanics as well as in quantum field theory~\cite{bonfiglioli2011topics, achilles2012early, tung1985group}.  

Although the BCH formula \eqref{BCH1} has been known for a long time, only few exact expressions have been found. The Heisenberg Lie algebra used in Quantum Mechanics is, perhaps, the most known example. In this case, the algebra is nilpotent \cite{dixmier1977enveloping} and (\ref{ExpConm}) becomes ${\rm e}^{L_X} Y = Y + [X,Y]$. Consequently (\ref{BCH1}) reduces to $Z(X,Y) = X+Y + \frac{1}{2}[X,Y]$. For more general Lie algebras,  \eqref{ExpConm} involves an infinite sum of nested commutators, making it difficult to derive an analytic expression for $Z(X,Y)$.

 During the last years several new results have been found and have generated a renewal of attention on this subject \cite{reinsch2000simple, casas2009efficient, blanes2009magnus, van2016simplifying}. In \cite{van2015special} Van-Brunt and Visser obtained an explicit formula for the case $[X,Y]=u X+v Y+c I$. Later on, Matone \cite{matone2015algorithm,matone2015classification} found an algorithm to extend such formula to the case where there are two commutators of the above form. Recently, Van-Brunt and Visser \cite{van2015explicit} generalized their previous result to the case where the Lie algebra satisfies some general requirements.   
 
  In this work we introduce a different Lie algebra, named \emph{contact Heisenberg Algebra} (CHA) and compute the exact closed form for the BCH formula (see equation (\ref{BCHfinal})). The CHA is the Lie algebra of linear functions over a contact manifold with the commutator given by the Lagrange bracket \cite{arnold2001dynamical}. This algebra is the analogue of the Heisenberg Lie algebra \cite{dixmier1977enveloping, moretti2013spectral} of linear functions over a symplectic manifold for the contact case.
  Moreover, it is a particular case of a Jacobi (or local Lie) algebra \cite{kirillov1976local}.
  
  The motivation for this work comes from the fact that contact manifolds naturally extend symplectic manifolds and are used in a wide range of physical theories.  For example, in mechanics \cite{bravetti2016contact,cariglia2016eisenhart,de2016geometric}, statistical mechanics~\cite{bravetti2015liouville,PhysRevE.93.022139} and thermodynamics \cite{mrugala1978geometrical,mrugala1991contact,rajeev2008hamilton,quevedo2007geometrothermodynamics,Bravetti2015377,bravetti2015sasakian} 
  (see also \cite{eberard2007extension,favache2010entropy,wang2015stabilization,kholodenko2013applications,grmela2014contact,ShinItiro1,ShinItiro2,martelli2006toric,Pan2015} for further applications). 
  Besides, some attempts in order to quantize contact systems have been proposed within the formalisms of geometric and deformation 
  quantization \cite{rajeev2008quantization, fitzpatrick2011geometric}. Our calculation of the BCH formula paves the way for a standard quantization program based on linear operators \cite{moretti2013spectral}.

	\section{The contact Heisenberg algebra}\label{ca}
Contact manifolds are the object of extensive research both in mathematics and physics. Hence, for a detailed introduction we refer the reader to e.g. references~\cite{arnold,arnold2001dynamical,geiges2008introduction}. The relevant aspect for our purposes is the following: using the geometric (contact) structure one can define a natural bracket of functions over a contact manifold. In the standard (Darboux) coordinates $(S,p_{i},q^{i})$, $i=1,2, \dots, n$, the bracket reads
	\beq\label{bracket}
	\{ g, f \} =
	\left( g \frac{\partial f}{\partial S} -  \frac{\partial g}{\partial S} f \right) + p_i \left(  \frac{\partial g}{\partial S} \frac{\partial f}{\partial p_i} -  \frac{\partial g}{\partial p_i} \frac{\partial f}{\partial S}  \right) + 					
	\left(  \frac{\partial g}{\partial q^i} \frac{\partial f}{\partial p_i} -  \frac{\partial g}{\partial p_i} \frac{\partial f}{\partial q^{i}} \right), 
	\eeq
where a sum over repeated indices is assumed here and below.
The bracket \eqref{bracket} is called the \emph{Lagrange bracket} \cite{arnold2001dynamical}.  
It differs from the Poisson bracket in some important properties. For instance, while it satisfies the Jacobi identity, 
 it fails to satisfy the Leibniz rule and thus it is not a derivation.  

Let us consider the space ${\cal S}$ of linear functions on the contact manifold, i.e. the space 
	\begin{equation} \label{ObsvS}
	{\cal S} = \mbox{Span}\{  z1 + a_{i} q^{i} + b^{i} p_{i} + c S \}, \qquad z,a_{i},b^{i},c \in \mathbb{R}\,.
	\end{equation}
By definition, ${\cal S}$ is a vector space with the natural sum operation. We can identify a generic linear function $f=z1 + a_{i} q^{i} + b^{i} p_{i} + c S$ with a vector $X_{f}$ 
and define a commutator between vectors in the following way
	\beq\label{commutator}
	[X_{f},X_{g}]=X_{\{f,g\}}\,.
	\eeq
It follows directly that ${\cal A} = ({\cal S}, +, [\,\cdot\,])$ is a (local) Lie algebra \cite{kirillov1976local}, 
that we call the \emph{contact Heisenberg algebra}. Note that this algebra is the analogue of the Heisenberg Lie algebra for the case of a contact manifold.
A natural basis for this algebra is $(X_{1},X_{q^{i}},-X_{p_{i}},X_{S})$, where the minus sign in $X_{p_{i}}$ is chosen for convenience to simplify the calculations.
 Using \eqref{bracket} and \eqref{commutator} we find that the only non-vanishing commutators
of the elements of the basis are
	\begin{equation}\label{commrelations}
	[ X_{1} , X_{S} ] = X_{1}, \qquad [X_{q^i}, X_{S} ] =  X_{q^i}, \qquad [X_{q^i} , X_{p_j} ] = \delta^i_jX_{1}\,.
	\end{equation}
	
In the following we assume for simplicity that the contact manifold is $3$-dimensional ($n=1$) so that the index on $q$ and $p$ can be omitted. The generalization to the $(2n+1)-$dimensional case is straightforward. Furthermore, it will be convenient to represent any vector $X_{f}$ as $X_{(z,a,b,c)}$, being $(z,a,b,c)$ its four components with respect to the basis $(X_{1},X_{q},-X_{p},X_{S})$.

Remarks:
 First, we remind that the subspace  ${\cal Z}({\mathfrak{g}})$ of elements $X \in  {\mathfrak{g}}$ such that $[X , Y] = 0, \; \forall\, Y \in {\mathfrak{g}}$ is called the \emph{center} (or \emph{centralizer}) of the algebra ${\mathfrak{g}}$ \cite{dixmier1977enveloping}. It is easy to check that for the contact Heisenberg algebra ${\cal A}$, the center ${\cal Z}({\cal A})$ is just the null vector $X_{(0,0,0,0)}$, i.e. it is trivial. Notice that this fact establishes a difference with both the Heisenberg Lie algebra and the generalization considered by Visser in \cite{van2015special}. In such cases the center is a 1-dimensional vector space.
 
 Secondly, the sub-algebra ${\tilde{\cal A}} \subset {\cal A}$ defined by ${\tilde{\cal A}} = \mbox{Span}\{ X_{(z,a,b,0)} \} $ is a (maximal) ideal of ${\cal A}$ and its center is the 1-dimensional vector space ${\cal Z}({\tilde{\cal A}}) = \mbox{Span}\{ X_{(z,0,0,0)}\}$.
It can be checked that the sub-algebra ${\tilde{\cal A}}$ is isomorphic to the Heisenberg Lie algebra, that is, the Heisenberg Lie algebra is naturally contained in ${\cal A}$.

{Finally, we stress that the study of
the BCH formula for the CHA does not correspond to any of the cases considered previously in the literature 
(see e.g. \cite{van2015explicit,matone2015algorithm,matone2015classification}) and therefore it is a completely new problem.}

\section{BCH for the contact Heisenberg algebra}\label{BCHca}

In this section we prove the main result of our work. We use the BCH formula \eqref{BCH1} and the commutation relations \eqref{commrelations} to find an exact formula for the element $Z(X,Y)$. As the CHA does not fall into any of the cases studied in previous works, we need to compute the infinite sum of nested commutators in \eqref{ExpConm} explicitly, which amounts to a rather lengthy calculation. Nevertheless, the final result has a very compact form. To simplify the presentation we split the calculation into three different steps.

\subsection{Step 1: calculation of ${\rm e}^{L_{X}}$}\label{subsec:exp}
The goal of this step is to obtain an expression for ${\rm e}^{L_{X_{(z,a,b,c)}} } X_{(\bar{z},\bar{a},\bar{b},\bar{c})}$ using the commutators \eqref{commrelations}. Let us start with the calculation of the first commutator, that is
	\begin{equation}
	L_{X_{(z,a,b,c)}} X_{(\bar{z},\bar{a},\bar{b},\bar{c})} = [ X_{(z,a,b,c)} , X_{(\bar{z},\bar{a},\bar{b},\bar{c})}]  =:  X_{(z_1 , a_1, 0, 0)}\,,
	\end{equation}
where we have defined
	\begin{equation}\label{eq1}
	z_1:={z \bar{c} - c \bar{z}} + {\bar{a} b - a \bar{b}}\,, \quad a_1:= {a \bar{c} - c \bar{a}}.
	\end{equation}
In the same way we can compute the second commutator
	\beq
	L_{X_{(z,a,b,c)}} L_{X_{(z,a,b,c)}} X_{(\bar{z},\bar{a},\bar{b},\bar{c})}  = 
	[ X_{(z,a,b,c)},  [ X_{(z,a,b,c)} , X_{(\bar{z},\bar{a},\bar{b} ,\bar{c})}]] 
	=[ X_{(z,a,b,c)}, X_{(z_1 , a_1, 0,  0)} ] =:  X_{(z_2, a_2, 0, 0)}\,,
	\eeq
where 
	\begin{equation}
	z_2:= - c\,z_1 +  a_1 b\,, \quad a_2:= - {c}\,a_1\,.
	\end{equation}
From the above relations, we can infer the higher order commutators. Following the same notation, we  write in general
	\begin{eqnarray}
	(L_{X_{(z,a,b,c)}})^n X_{(\bar{z},\bar{a},\bar{b},\bar{c})} = X_{(z_n, a_n, 0, 0)}\,,
	\end{eqnarray}
 being
	\begin{equation}
	z_n:= - {c}\, z_{n-1} + a_{n-1} b , \quad a_n:= -c\, a_{n-1}\,.
	\end{equation}
Thus we obtain a recursive relation that gives $z_n$ and $a_n$ in terms of $z_{n-1}$ and $a_{n-1}$. Proceeding by induction one can show that such relations imply the following
	\begin{equation}\label{anbn}
	a_n = \left(-{c}\right)^{n-1} a_{1}, \qquad  z_{n} = \left(- c\right)^{n-1} z_{1} + \left( - c\right)^{n-2}  {(n-1) a_1 \,  b }\,.
	\end{equation}
Now we are ready to compute the action of ${\rm e}^{L_X}$. Let us rewrite explicitly the exponential as
        \begin{eqnarray}
        {\rm e}^{L_{X_{(z,a,b,c)}} } X_{(\bar{z},\bar{a},\bar{b},\bar{c})} &=& \left[{I} +  L_{X_{(z,a,b,c)}}  + \frac{1}{2!} (L_{X_{(z,a,b,c)}} )^2 + \cdots + \frac{1}{n!} (L_{X_{(z,a,b,c)}} )^n + \dots \right] X_{(\bar{z},\bar{a},\bar{b},\bar{c})} 
         \\
        &=& X_{(\bar{z},\bar{a},\bar{b},\bar{c})} + X_{(z_1, a_1, 0, 0)} + \frac{1}{2!} X_{(z_2, a_2, 0, 0)} + \cdots + \frac{1}{n!} X_{(z_n, a_n, 0, 0)} + \dots \nonumber 
        \end{eqnarray}
Next, we decompose each operator in the Lie algebra basis $(X_{1},X_{q},-X_{p},X_{S})$ and sum similar terms to obtain
	\begin{equation}
	 {\rm e}^{L_{X_{(z,a,b,c)}} } X_{(\bar{z},\bar{a},\bar{b},\bar{c})}= 
	\left( \bar{z} + z_1 + \frac{ z_2}{2!}  + \dots + \frac{ z_n}{n!}  + \dots \right) X_{1} + \left( \bar{a} + a_1 + \frac{a_{2}}{2!} + \dots + \frac{a_{n}}{n!} + \dots \right)X_{q} - \bar{b} X_{p} + \bar{c}\, X_{S}\,,
	\end{equation}
 where $a_{i}$ and $z_{i}$ are as in \eqref{anbn}. First, let us consider the $X_{q}$ component in the above formula. We write it as
        \begin{eqnarray}
        a_{\infty} & := &
        \bar{a} + a_1 + \frac{a_{2}}{2!} + \dots + \frac{a_{n}}{n!}+ \dots = \bar{a} + a_1 \left[ 1 + \frac{1}{2!}  \left( - {c} \right) + \dots + \frac{1}{n!} \left( - {c} \right)^{n-1} + \dots \right] \nonumber \\
        & =& \bar{a} - \frac{ a_1}{c} \left[ \left( - {c} \right) + \frac{1}{2!}  \left( - {c} \right)^2 + \dots + \frac{1}{n!} \left( - {c} \right)^{n} + \dots \right] = \bar{a}\, {\rm e}^{-{c}} + \frac{a \bar{c}}{c} (1-{\rm e}^{-c})\label{Xqcomponent}\,.
        \end{eqnarray}

For as regards the $X_{1}$ component, we have
	\begin{eqnarray}\label{X1component}
	z_{\infty} &=& \bar{z} + z_1 + \frac{z_{2}}{2!} + \dots + \frac{z_{n}}{n!} + \dots \nonumber \\
	&=& \bar{z} + z_1 + \frac{1}{2!} \left[ - {c}\, z_{1} + {a_1}b \right]+\frac{1}{3!} \left[ \left( - {c}\right)^2 z_1 - 2 {c}{a_1 b } \right]  + \dots + \frac{1}{n!} \left[  \left( -{c}\right)^{n-1} z_1 + (n-1) \left( - c\right)^{n-2}  a_1 b\right] + \dots 
	\end{eqnarray}
We notice that in the above formula we can collect $\bar z$ and the terms in $z_{1}$ to obtain 
	\begin{eqnarray}\label{z1terms}
	\bar z + z_1 + \frac{1}{2!} \left( - {c}\right) z_1 + \dots + \frac{1}{n!} \left( -{c}\right)^{n-1} z_1 + \dots = \bar{z} + \frac{z_1}{c} ( 1 - {\rm e}^{-c})  \,.
	\end{eqnarray}
Now let us consider the remaining terms in \eqref{X1component}, which can be rearranged as
	\beq\label{finalcoeff}
	  \frac{a_1 b}{c^2} \left[ \frac{1}{2!} \left( -c\right)^2 + \frac{2}{3!}\left( -{c}\right)^3 + \dots + \frac{(n-1)}{n!} \left( -{c}\right)^{n} + \dots  \right] = \frac{a_1 b}{c^2} \left[ 1 - {c}\, {\rm e}^{-c} - {\rm e}^{-c} \right]\,,
	\eeq
where we made use of the identity
	\begin{equation}\label{usefulformula}
	\frac{1}{2!} \alpha^2 + \frac{2}{3!} \alpha^3 + \dots + \frac{(n-1)}{n!} \alpha^n + \dots = 1 + \alpha\, {\rm e}^\alpha - {\rm e}^\alpha\,.
	\end{equation}
Using \eqref{X1component}, \eqref{z1terms} and \eqref{finalcoeff} we finally get the expression for the $X_{1}$ component, which reads
	\begin{eqnarray}\label{zinfinity}
	 z_{\infty} &=&  \bar{z} + \frac{z_1}{c} ( 1 - {\rm e}^{-c})  + \frac{a_1 b}{c^2} \left( 1 -{c}\, {\rm e}^{-{c}} - {\rm e}^{-{c}} \right) \nonumber \\
	 &=& \bar{z}\, {\rm e}^{-c} + \bar{a} b {\rm e}^{-c} - \frac{a  \bar{b}}{c} (1 - {\rm e}^{-c} ) + \frac{\bar{c}}{c} \left[ z (1 - {\rm e}^{-c} )  + \frac{ab}{c} (1 - c {\rm e}^{-c} -{\rm e}^{-c} )  \right] \, ,
	 	\end{eqnarray}
\noindent where the last equality follows by substituting $a_{1}$ and $z_{1}$ from \eqref{eq1}.

Summing up, we were able to find out an analytic expression for the infinite series appearing in $ {\rm e}^{L_{X_{(z,a,b,c)}} }$. As a result the exponential operator acts as
	\begin{eqnarray}\label{expfinal}
	 {\rm e}^{L_{X_{(z,a,b,c)}} } X_{(\bar{z},\bar{a},\bar{b},\bar{c})} &=& z_{\infty} X_{1} + a_{\infty}X_{q} - \bar{b}X_{p} +\bar{c}X_{S}=X_{(z_{\infty},a_{\infty},\bar{b},\bar{c})}\,,
	\end{eqnarray}
with $a_{\infty}$ and $z_{\infty}$ given in equations \eqref{Xqcomponent} and \eqref{zinfinity} respectively.

Since this is a linear operator, we can represent it as a matrix using the basis $(X_{1},X_{q},-X_{p},X_{S})$ as follows
	\begin{equation}\label{matrixrepr}
	 {\rm e}^{L_{X_{(z,a,b,c)}} } X_{(\bar{z},\bar{a},\bar{b},\bar{c})} :={M}(z,a,b,c) \left( \begin{array}{c} \bar{z} \\ \bar{a} \\ \bar{b} \\ \bar{c} \end{array} \right)
	= \left( \begin{array}{cccc}  {\rm e}^{-c} &  b  {\rm e}^{-c} & -\frac{a}{c} (1- {\rm e}^{-c}) & \frac{z}{c} \left( 1 - {\rm e}^{-c}\right) + \frac{ab}{c^2} (1 - c {\rm e}^{-c} - {\rm e}^{-c} )   \\ 0 &  {\rm e}^{-c} & 0 & \frac{a}{c} (1  -{\rm e}^{-c}) \\
	0 & 0 &  1  & 0 \\ 0 & 0 & 0 & 1 \end{array}\right) \left( \begin{array}{c} \bar{z} \\ \bar{a} \\ \bar{b} \\ \bar{c} \end{array} \right)\,.
	\end{equation}

\subsection{Step 2 : calculation of $(I-{\rm e}^{L_{X}}{\rm e}^{t L_{Y}})^{n}$}\label{subsec:Iexp}
Taking advantage of the matrix representation \eqref{matrixrepr}, we can rewrite {the product of the two exponentials involved in (\ref{BCH1}) as}
	\beq\label{expmatrix}
	{\rm e}^{L_{ X_{(z,a,b,c)}} } {\rm e}^{ t L_{X_{(\bar{z}, \bar{a}, \bar{b}, \bar{c})}} }={M}(z,a,b,c) {M}(t \bar{z},t \bar{a}, t \bar{b}, t \bar{c}) =: M\,.
	\eeq
Using the structure of the matrix in \eqref{matrixrepr}, it is easy to check that $M$ (and any of its powers) retains the same form
	\begin{equation}
	M = \left( \begin{array}{cccc} {\rm e}^{m} & M_{12} & M_{13} & M_{14} \\ 0 & {\rm e}^{m} & 0 & M_{24} \\ 0 & 0 & 1 & 0 \\ 0 & 0 & 0 & 1 \end{array}\right)\,, \label{DefM}
	\end{equation}
where
	\begin{eqnarray}\label{mform}
	m&=&-c-t\bar{c}\,, \quad M_{12}= {\rm e}^{m}(b+t\bar{b})\,, \quad M_{13}=- \frac{a}{c} (1- {\rm e}^{-c}) - \frac{\bar{a}}{\bar{c}} {\rm e}^{-c} (1 - {\rm e}^{-t \bar{c}} ), \quad M_{24}= \frac{a}{c}(1-{\rm e}^{-c})+\frac{\bar{a}}{\bar{c}}({\rm e}^{-c}-{\rm e}^{m}),
	 \nonumber\\
	M_{14}&=& \frac{\bar{z}}{\bar{c}} (1- {\rm e}^{- t \bar{c}}){\rm e}^{-c} + \frac{z}{c} (1- {\rm e}^{- c})  + \frac{\bar{a} \bar{b}}{\bar{c}^2} \left( 1 - t \bar{c} {\rm e}^{- t \bar{c}} - {\rm e}^{- t \bar{c}} \right){\rm e}^{-c}  + \frac{\bar{a} b}{\bar{c}} (1 - {\rm e}^{- t \bar{c}}) {\rm e}^{-c}  + \frac{ab}{c^2} (1 - c {\rm e}^{-c} - {\rm e}^{-c} ) .
	\end{eqnarray}

Now we use these properties of $M$ to write $(I-{\rm e}^{L_{ X_{(z,a,b,c)}} } {\rm e}^{ t L_{X_{(\bar{z}, \bar{a}, \bar{b}, \bar{c})}} })^{n}$ in a compact form
	\begin{eqnarray}\label{imenosmalan}
	(I-{\rm e}^{L_{ X_{(z,a,b,c)}} } {\rm e}^{ t L_{X_{(\bar{z}, \bar{a}, \bar{b}, \bar{c})}} })^{n}&=&(I-M)^{n}=\sum_{k=0}^{n}\frac{n!(-1)^{k}}{(n-k)!k!}M^{k}\,,
	\end{eqnarray}
from which we observe that we are interested in computing the powers of the matrix $M$. After some algebra, it can be shown that 
	\begin{equation}\label{Mk}
	M^k = \left( \begin{array}{cccc} {\rm e}^{km} & k {\rm e}^{(k-1)m} M_{12} &  \left( \frac{1-{\rm e}^{km}}{1-{\rm e}^{m}}\right) M_{13} & \left(M^{k}\right)_{14} \\ 0 & {\rm e}^{km} & 0  &  \left( \frac{1-{\rm e}^{km}}{1-{\rm e}^{m}}\right)M_{24}  \\ 0 & 0 & 1
	& 0 \\ 0 & 0 & 0 & 1\end{array} \right)\, ,
	\end{equation}
where
	\beq\label{Mk14}
	(M^{k})_{14}= \frac{(1-{\rm e}^{km})}{(1-{\rm e}^m)} M_{14} + \frac{[1-k{\rm e}^{(k-1)m}+(k-1){\rm e}^{km}]}{(1-{\rm e}^m)^2} M_{12} M_{24} \,.
	\eeq
Finally, \eqref{imenosmalan} and \eqref{Mk} imply that
	\begin{eqnarray} \label{PoliM}
	(I-M)^{n}=\left( \begin{array}{cccc} (1-{\rm e}^{m})^n &  - n (1-{\rm e}^m)^{n-1} M_{12} &   -  (1-{\rm e}^m)^{n-1} M_{13} & {\cal M} \\ 0 & (1-{\rm e}^{m})^n 
	 & 0  & - (1-{\rm e}^{m})^{n-1} M_{24} \\ 0 & 0 & 0 & 0 \\ 0 & 0 & 0 & 0 \end{array}\right),
	\end{eqnarray}
where 
	\beq
	{\cal M}=\sum^n_{k=0} \frac{n! (-1)^k}{(n-k)! \, k!} (M^k)_{14} = - (1-{\rm e}^m)^{n-1} M_{14}  + (n-1)(1-{\rm e}^m)^{n-2} M_{12} M_{24} \,. \label{Mgrande}
	\eeq

This result facilitates the calculation of the series inside the integral in the BCH formula, which will be done in the next step.  
	
\subsection{Step 3: calculation of the BCH formula}\label{subsec:BCH}

This last step is devoted to the calculation of the exact BCH formula for the CHA. Notice that using our notation we can rewrite \eqref{BCH1} as
	\begin{equation}\label{BCH2}
	Z(X_{(z,a,b,c)}, X_{(\bar{z}, \bar{a}, \bar{b}, \bar{c})}) = X_{(z,a,b,c)} + X_{(\bar{z}, \bar{a}, \bar{b}, \bar{c})} -  \int^1_0 dt \, \sum^{+\infty}_{n=1} \frac{\left(I-M\right)^n}{n(n+1)}  X_{(\bar{z}, \bar{a}, \bar{b}, \bar{c})}.
	\end{equation}
By considering (\ref{PoliM}) and (\ref{Mgrande}) we compute the term inside the integral which takes the form
	\beq\label{finalsum}
	\sum^{+\infty}_{n=1} \dfrac{\left(I-M\right)^n}{n(n+1)} = \left( \begin{array}{cccc} \frac{1-{\rm e}^m + m{\rm e}^m}{1-{\rm e}^{m}} & \frac{M_{12} (1+m-{\rm e}^m)}{(1-{\rm e}^m)^2} &  - \frac{M_{13} (1-{\rm e}^m + m{\rm e}^m)}{(1-{\rm e}^{m})^2}  & m_{14}  \\ 0 & \frac{1-{\rm e}^m + m{\rm e}^m}{1-{\rm e}^{m}} & 0  & - \frac{M_{24} (1-{\rm e}^m + m{\rm e}^m)}{(1-{\rm e}^{m})^2} \\ 0 & 0 & 0  & 0 \\ 0 & 0 & 0 & 0 \end{array} \right) =: \tilde{M}\,,
	\eeq
where 
	\beq\label{m14}
	m_{14} := - \frac{(1-{\rm e}^m+m{\rm e}^m) M_{14}}{(1-{\rm e}^m)^2} - \frac{[2(1-{\rm e}^{m}) +m(1+{\rm e}^m)]  M_{12} M_{24} }{(1-{\rm e}^m)^3}\,.
	\eeq

Finally, we integrate (\ref{finalsum}) and obtain
	\beq\label{integral1}
	\int^1_0 dt \, \tilde{M} X_{(\bar{z}, \bar{a}, \bar{b}, \bar{c})} = 
	\left( \begin{array}{c} f(c,\bar{c}) \left( \bar{c} z- \bar{z} c + \frac{\bar{c}}{c} ab - \frac{c}{\bar{c}} \bar{a} \bar{b} \right) + \left( g_1(c,\bar{c}) \frac{b}{c} - g_2(c,\bar{c}) \frac{\bar{b}}{\bar{c}} \right) \left( \bar{a} c- \bar{c} a \right)   \\  
	 f(c,\bar{c}) \left( \bar{c} a- \bar{a} c \right) \\ 
	 0 \\ 0 \end{array} 
	\right)\,,
	\eeq
with
	\begin{eqnarray}
	f(c,\bar{c}) &:=& \frac{1}{\left( 1 - {\rm e}^{c + \bar{c}} \right)} \left[  \frac{\left( 1 - {\rm e}^{ \bar{c}} \right)}{\bar{c}} - \frac{{\rm e}^{\bar{c}} \left( 1 - {\rm e}^{c} \right)}{c} \right] \,, \\
	g_1(c,\bar{c}) &:=& \frac{\left(1 - {\rm e}^{\bar{c}} \right)}{\bar{c}} \left[ \frac{ 1}{ (1 - {\rm e}^{c + \bar{c}})} + \frac{( c + \bar{c}) {\rm e}^{c + \bar{c}} }{ (1 - {\rm e}^{c + \bar{c}})^2}  \right]  \, , \\
	g_2(c,\bar{c}) &:=& {\rm e}^{\bar{c}} \frac{\left(1 - {\rm e}^{c} \right)}{c} \left[ \frac{ 1}{ (1 - {\rm e}^{c + \bar{c}})} + \frac{( c + \bar{c}) }{ (1 - {\rm e}^{c + \bar{c}})^2}  \right]  \, 
	\end{eqnarray}
\noindent Using these relations, we arrive at our main result, the exact expression of the BCH formula \eqref{BCH1} for the CHA, which reads
	\begin{eqnarray}
	Z(X_{(z,a,b,c)}, X_{(\bar{z}, \bar{a}, \bar{b}, \bar{c})}) &=& X_{(z,a,b,c)} + X_{(\bar{z}, \bar{a}, \bar{b}, \bar{c})}-f(c,\bar{c})\left[X_{(z,a,b,c)}, X_{(\bar{z}, \bar{a}, \bar{b}, \bar{c})}\right]+\nonumber\\
	&+& \left( \bar{a} c - a \bar{c} \right) \left[ \left(f(c,\bar{c})-g_{1}(c,\bar{c})\right) \frac{b}{c} + \left(f(c,\bar{c})+g_{2}(c,\bar{c})\right) \frac{ \bar{b}}{\bar{c}} \right]X_{(1,0,0,0)}\,.\label{BCHfinal}
	\end{eqnarray}
Formula (\ref{BCHfinal}) provides the multiplication of the Lie group associated to the CHA.
We notice that  \eqref{BCHfinal} is similar in form to the results obtained by Van-Brunt and Visser and by Matone \cite{van2015special, matone2015algorithm, matone2015classification, van2015explicit}. 
However, in our case there is an additional term, which is proportional to the generator $X_{(1,0,0,0)}$. Moreover, the functions $f(c,\bar c)$ and $g_{i}(c,\bar c)$ are different from those in
\cite{van2015special, matone2015algorithm, matone2015classification, van2015explicit}.

Recall that the parameters $c$ and $\bar{c}$ characterize the (contact) component $X_{S}$ of the elements of the CHA (see \eqref{ObsvS}). 
Furthermore, the sub-algebra $\tilde{\cal A}$ of elements of the type $X_{(z,a,b,0)}$ is isomorphic to the Heisenberg Lie algebra  (see the remarks in section \ref{ca}).
The same behavior can be expected at the level of the corresponding Lie groups.
In fact, in the limit $c,\bar{c}\rightarrow 0$ 
the expression \eqref{BCHfinal} reduces to
	\beq\label{Hlimit}
	\lim_{c,\bar c \rightarrow 0}Z(X_{(z,a,b,c)}, X_{(\bar{z}, \bar{a}, \bar{b}, \bar{c})}) = X_{(z,a,b,c)} + X_{(\bar{z}, \bar{a}, \bar{b}, \bar{c})}+\frac{1}{2}\left[X_{(z,a,b,c)}, X_{(\bar{z}, \bar{a}, \bar{b}, \bar{c})}\right]
	=:Z_{H}(X_{(z,a,b,0)}, X_{(\bar{z}, \bar{a}, \bar{b},0)})\,, 
	\eeq
where $Z_{H}$ is the known BCH formula for the Heisenberg Lie algebra.
To first order in $c$ and $\bar c$ we have the correction to $Z_{H}$, given as
	\beq \label{DefMulti}
	Z(X_{(z,a,b,c)}, X_{(\bar{z}, \bar{a}, \bar{b}, \bar{c})}) = Z_{H}(X_{(z,a,b,0)}, X_{(\bar{z}, \bar{a}, \bar{b},0)})-
	\frac{(c-\bar c)}{12}\left[X_{(z,a,b,c)}, X_{(\bar{z}, \bar{a}, \bar{b}, \bar{c})}\right]-\frac{(\bar a c-a \bar c)(b-\bar b)}{12}\,X_{(1,0,0,0)}+\dots
	\eeq
	
	This relation can be used to explore the linear deformation of the Heisenberg group coming from the Lie group associated to the CHA.

\section{Conclusions}

In this work we provide the exact expression of the BCH formula for the contact Heisenberg algebra (CHA), see (\ref{BCHfinal}), the analogue of the Heisenberg Lie algebra for contact manifolds. 
This formula gives the multiplication of the Lie group associated to the CHA.

Similarly to previous works \cite{van2015special, matone2015algorithm, matone2015classification, van2015explicit}, our result depends only up to the first commutator. 
However, there is an additional term proportional to the constant function $X_{(1,0,0,0)}$. Moreover, the functions $f(c,\bar c)$ and $g_{i}(c,\bar c)$ are different from those in
\cite{van2015special, matone2015algorithm, matone2015classification, van2015explicit}. 

Remarkably, in the limit $c, \bar{c} \rightarrow 0$, we recover the BCH formula for the Heisenberg Lie algebra. This result shows that in the same way as the Heisenberg Lie algebra is contained in the CHA, the Heisenberg group can be seen as a subgroup of the Lie group associated to the CHA. Additionally, we calculate the linear deformations of the Heisenberg group (\ref{DefMulti}). 

In the standard quantization program \cite{moretti2013spectral}, the Heisenberg group is promoted to an algebra  (the Weyl algebra) and its elements are represented (via the Gelfand-Naimark-Segal construction) as operators in a Hilbert space. We plan to follow the same route in the contact case. The present calculation of the BCH formula constitutes the first step in this direction.

Finally, as the CHA is a particular case of a Jacobi (or local Lie) algebra~\cite{kirillov1976local}, 
we believe that our calculations can be useful to obtain a BCH formula for more general local Lie algebras.

\section*{Acknowledgements}
The authors would like to thank Jos\'e Figueroa-O'Farrill for his relevant comments on the first version of this work. AB is supported by a DGAPA-UNAM postdoctoral fellowship. Angel Garcia-Chung acknowledges the total support from DGAPA-UNAM fellowship and partial support from CONACYT project 237503 and DGAPA-UNAM grant IN 103716. DT acknowledges financial support from CONACYT, CVU No. 442828.

\bibliographystyle{ieeetr}
\bibliography{Conservative}


\end{document}